\documentclass{ieeeaccess}
\usepackage{cite}
\usepackage{amsmath,amssymb,amsfonts}
\usepackage{algorithmic}
\usepackage{graphicx}
\usepackage{textcomp}
\usepackage{algorithm}
\usepackage{algorithmic}
\usepackage{seqsplit}
\usepackage{etoolbox}
\makeatletter
\providecommand{\xfigwd}{\the\linewidth}

\patchcmd{\@makecaption}
  {\ifdim \xfigwd >\columnwidth}
  {\ifdim \xfigwd >\textwidth}
  {}{}
\makeatother
\usepackage{etoolbox}
\makeatletter
\let\orig@fontseries\fontseries
\renewcommand{\fontseries}[1]{%
  \ifx#1n%
    \orig@fontseries{m}% 
  \else
    \orig@fontseries{#1}%
  \fi}
\makeatother

\usepackage[colorlinks,urlcolor=blue,linkcolor=blue,citecolor=blue]{hyperref}
\expandafter\def\expandafter\UrlBreaks\expandafter{\UrlBreaks\do\/\do\*\do\-\do\~\do\'\do\"\do\-}
\usepackage{color}
\usepackage{tabularx}
\usepackage{booktabs}
\usepackage{siunitx}
\usepackage{makecell}
\usepackage{cleveref}

\DeclareRobustCommand{\circuit}[1]{\texttt{\expandafter\seqsplit\expandafter{\detokenize{#1}}}}

\def\BibTeX{{\rm B\kern-.05em{\sc i\kern-.025em b}\kern-.08em
    T\kern-.1667em\lower.7ex\hbox{E}\kern-.125emX}}
\begin{document}
\history{}
\doi{}

\title{A Neutral-Atom Quantum Compiler with Application-Specific Layout and Hub-Assisted Shuttling}

\author{
    \uppercase{Takahiko Satoh}\authorrefmark{1}, 
    and
    \uppercase{Takaharu Yoshida}\authorrefmark{2},
}
\address[1]{Faculty of Science and Technology, Keio University, Yokohama, Japan (email: satoh@ics.keio.ac.jp)}
\address[2]{Department of Physics, Graduate School of Science, Tokyo University of Science, Tokyo, Japan (email: 1225707@ed.tus.ac.jp)}
\tfootnote{
This paper is based on results obtained from a project,
JPNP23003, commissioned by the New Energy and Industrial Technology Development Organization (NEDO).
TS is also supported by JST Grant Number JPMJPF2221.
}
\markboth
{Satoh \headeretal: A Neutral-Atom Quantum Compiler with Application-Specific Layout and Hub-Assisted Shuttling}
{Satoh \headeretal: A Neutral-Atom Quantum Compiler with Application-Specific Layout and Hub-Assisted Shuttling}
\corresp{Corresponding author: Takahiko Satoh (email: satoh@ics.keio.ac.jp).}

\begin{abstract}
Compiling arbitrary-connectivity NISQ circuits onto monolithic single-zone neutral-atom devices is constrained by a finite interaction range and a minimum separation between simultaneously addressable sites.
Under the minimum-separation constraint, the SWAP-only configuration of our pipeline does not return a schedule within a practical time budget on a range of circuits, including circuits as small as nine qubits.
We address this with hub traps, a small number of dynamically placed empty traps that serve as transit waypoints, together with a per-gate rule that chooses between SWAP-based routing and hub-mediated shuttling.
We evaluate the compiler on seventeen benchmarks using analytic estimates of execution time and a per-layer fidelity proxy, comparing against a placement-matched baseline and against ablations of our own pipeline.
Hub traps make these otherwise-unsolved circuits compile in seconds to minutes and remove SWAP gates entirely on every completed circuit, so their role is to enable routing rather than only to optimize fidelity.
The benefit is concentrated on routing-dominated circuits and is absent on routing-free ones, which we separate by the structure of the interaction graph.
On the most routing-dominated circuit the fidelity proxy improves by up to three orders of magnitude over the placement-matched baseline. The gain comes primarily from eliminating SWAP overhead, as the absolute fidelities there remain low.
\end{abstract}

\begin{IEEEkeywords}
Neutral-atom quantum computing, quantum compilation, qubit placement, atom shuttling, hub traps, Rydberg blockade.
\end{IEEEkeywords}

\titlepgskip=-15pt

\maketitle
\section{Introduction}
\label{sec:introduction}

Neutral-atom quantum computing has rapidly advanced as a promising platform for scalable quantum processors.
Optical tweezer arrays provide flexible qubit layouts~\cite{barredo2016atom}, long coherence times, and Rydberg-mediated entangling operations.
The circuits of interest belong to the noisy intermediate-scale quantum (NISQ) regime~\cite{preskill2018nisq}, where limited depth and gate fidelity make the overhead of realizing nonlocal interactions a first-order concern.
At the same time, the physics and control stack introduce compilation challenges that differ fundamentally from
static-coupling architectures: entangling feasibility depends on geometry, parallelism is constrained by blockade,
and qubits can be physically transported during program execution~\cite{bluvstein2022quantum}.

These features create a tension.
On the one hand, application-specific placement can greatly reduce routing overhead by bringing frequently interacting qubits closer.
On the other hand, placement alone is often insufficient: nonlocal interactions still arise, and a compiler must decide how to
realize them under hardware constraints.
Two major directions have emerged.
Placement-centric approaches exploit layout flexibility to mitigate blockade effects and reduce pulses~\cite{patel2023graphine,silver2024qompose}.
Movement-centric approaches schedule atom transport to reduce or eliminate SWAP gates~\cite{ludmir2024parallax,wang2024atomique,tan2024dpqa}.
However, bridging these directions under a conservative hardware model remains nontrivial,
especially when additional geometric constraints from control optics and occupancy management are taken into account.

In this work, we propose a practical two-step compilation workflow for neutral-atom devices.
First, we perform application-specific placement by solving a continuous optimization problem that reduces long-distance interactions while enforcing a minimum-separation constraint.
Second, we introduce hub traps as temporary capacity to make shuttling feasible, and we transpile the circuit by dynamically choosing between SWAP-based routing and shuttling-based routing using a unified cost model.
This decomposition yields a compiler structure that is modular and tunable, while staying faithful to geometry-dependent constraints.

We evaluate the proposed workflow on seventeen benchmarks against a placement-matched baseline and against ablations of our own pipeline, using proxy estimates of execution time and success.
Two findings stand out.
First, under the minimum-separation constraint, the SWAP-only configuration of our pipeline fails to return a schedule within a practical time budget on a range of circuits, including small ones, while hub traps make the same circuits compile in seconds to minutes.
We therefore treat hub traps as an enabling capability rather than a fidelity optimization alone.
Second, the benefit is regime-dependent.
It is concentrated on circuits whose interaction graph forces long-range routing, and is absent on circuits that placement alone already resolves.
We characterize this separation and report the cases where the current method does not succeed.
\section{Related Work on Neutral-Atom Compilation}
\label{sec:related}

Neutral-atom quantum computers provide a distinct compilation landscape:
(i) two-qubit entangling operations are enabled by Rydberg interactions
within a finite interaction range,
(ii) parallel entangling gates are constrained by the Rydberg blockade,
and (iii) atoms can be physically transported (shuttled) during program execution.
These features motivate compiler designs that go beyond static coupling graphs.
On static-coupling devices such as superconducting processors, the corresponding problem is qubit mapping and routing, where a circuit is made executable by inserting SWAP gates that bring interacting qubits onto adjacent couplers, handled in practice by heuristic routers such as SABRE~\cite{li2019sabre}.
The underlying qubit-movement subproblem has also been studied formally as permutation routing on the architecture graph~\cite{childs2019circuit}.
Neutral-atom platforms reshape this problem, because the layout is application-specific and atoms can be physically repositioned, so routing need not rely on SWAP insertion alone.
\subsection{Hardware constraints shaping compilation objectives}
\label{sec:related:constraints}

A typical neutral-atom device arranges atoms in optical tweezers.
A two-qubit gate (e.g., CZ) can be executed only when the inter-atomic distance
is within an effective blockade/interaction radius \(r_b\).
This radius arises from the Rydberg blockade, the mechanism that underlies neutral-atom entangling gates and is reviewed in~\cite{saffman2010rydberg}, and high-fidelity controlled-phase gates of this kind have been demonstrated experimentally on neutral atoms~\cite{levine2019highfidelity}.
Moreover, when an entangling operation is applied to a pair, nearby atoms may be
blockaded, restricting which other entangling operations can be executed concurrently.
These constraints couple \emph{mapping}, \emph{scheduling}, and \emph{routing}.

Importantly, experimental operating points are often specified by a set of coupled
parameters, such as \((r_b, t_{\mathrm{CZ}}, F_{\mathrm{CZ}})\).
For example, changing the Rydberg excitation condition to speed up entangling gates
may also affect the effective interaction range and crosstalk, and vice versa.
Accordingly, many compilation studies assume a fixed operating point and evaluate
trade-offs under that assumption~\cite{henriet2020neutral,wintersperger2023neutral}.

Besides the interaction radius, optical addressing and control hardware impose additional
geometric restrictions. In practice, tightly focused beams have finite waist and steering
resolution, which may translate into a minimum separable distance among simultaneously
addressable sites. In this work, we explicitly include such a constraint as a minimum
distance \(d_{\min}\) among traps/sites (see Sec.~\ref{sec:method:model}), to avoid
geometries that would require unrealistically dense addressing.

\subsection{Neutral-atom compilation frameworks}
\label{sec:related:frameworks}

Several compiler frameworks target neutral-atom platforms from different angles.
\emph{Geyser} leverages native multi-qubit gates and pulse-level structure to reduce
laser pulse counts for neutral-atom systems~\cite{patel2022geyser}.
\emph{Atomique} targets reconfigurable atom arrays (field-programmable qubit arrays),
integrating qubit mapping, atom movement, and gate scheduling under hardware constraints~\cite{wang2024atomique}.
Tan \emph{et al.} study compilation for dynamically field-programmable qubit arrays (DPQA/FPQA),
highlighting the algorithmic opportunities and constraints induced by coherent atom movement~\cite{tan2024dpqa}.
Recent works continue to expand the design space with zoning/movement-aware compilation,
e.g., \emph{PowerMove}~\cite{ruan2024powermove}, and techniques focusing on execution schemes and
layout selection~\cite{patel2024qpilot,silver2024qompose,huang2025dasatom}.

\subsection{Placement-centric compilation}
\label{sec:related:placement}

A line of work emphasizes that neutral-atom platforms can realize \emph{application-specific}
two-dimensional layouts, and that careful placement can reduce both blockade conflicts and
routing overhead. \emph{Graphine} proposes algorithm-aware layouts and reports improvements in
gate/pulse metrics by exploiting layout flexibility~\cite{patel2023graphine}.
\emph{Qompose} further studies selecting algorithm-specific layouts to improve parallelism and fidelity
across benchmark families~\cite{silver2024qompose}.
These approaches motivate our first stage (Sec.~\ref{sec:method:placement}), which aims at
constructing an initial placement that reduces long-distance entangling demands while respecting
geometric feasibility.

\subsection{Shuttling- and movement-centric compilation}
\label{sec:related:movement}

Another line of work focuses on \emph{atom transport} as a routing primitive.
\emph{Parallax} proposes a zero-SWAP compilation method that schedules atom movements and gates in
a scalable, parallelizable way under hardware constraints~\cite{ludmir2024parallax}.
More recently, divide-and-shuttle style transformations (e.g., \emph{DasAtom}) illustrate how
movement can be used to transition between mappings across circuit partitions~\cite{huang2025dasatom}.
These works reinforce the importance of modeling movement cost and feasibility, which becomes
central in our second stage (Sec.~\ref{sec:method:transpilation}).

Q-Pilot~\cite{patel2024qpilot} is the closest design to ours in that it also routes long-range interactions on a reconfigurable atom array, but the two works rest on different hardware assumptions.
Q-Pilot maps data qubits to fixed SLM traps and routes two-qubit gates using movable ancilla atoms, called flying ancillas, that are generated and recycled during execution and require no atom transfer of the data qubits themselves.
Entangling gates are induced by a zone-wide global Rydberg laser.
Our setting differs on the three axes summarized in Table~\ref{tab:qpilot_comparison}.
We move main qubits through hub traps rather than introducing ancillas, we treat all atoms as equivalent rather than partitioning them into data and ancilla roles, and we assume per-atom local addressing rather than a zone-wide laser.
These differences place the two works in complementary regions of the design space rather than in direct competition.
\begin{table}[t]
\centering
\caption{Comparison of Q-Pilot~\cite{patel2024qpilot} and this work along three axes that distinguish the underlying hardware assumptions.}
\label{tab:qpilot_comparison}
\footnotesize
\setlength{\tabcolsep}{4pt}
\renewcommand{\arraystretch}{1.15}
\begin{tabularx}{\linewidth}{>{\raggedright\arraybackslash}p{0.26\linewidth} >{\raggedright\arraybackslash}X >{\raggedright\arraybackslash}X}
\toprule
Axis & Q-Pilot & This work \\
\midrule
Main-qubit mobility & Data qubits fixed in SLM traps; only ancillas move & Main qubits themselves transit through hub traps \\
Ancilla model & Movable AOD atoms act as flying ancillas, generated and recycled during execution & No ancilla concept; all atoms are equivalent, hubs are transient empty traps \\
Rydberg addressing & Zone-wide global laser activating all atoms & Per-atom local addressing \\
\bottomrule
\end{tabularx}
\end{table}

\subsection{Positioning of this work}
\label{sec:related:positioning}

Our work is motivated by a practical gap between (i) placement-centric compilation that assumes
that a good initial layout is sufficient, and (ii) movement-centric compilation that heavily relies
on shuttling while often abstracting away additional geometric constraints arising from control optics.
We propose a two-step approach:
\emph{application-specific placement} followed by \emph{hub-assisted movement optimization}, where the
latter includes an explicit SWAP-versus-shuttling decision with a unified cost model.
This design aims to retain the strengths of both directions while maintaining a conservative and
configurable hardware model.

\section{Proposed Method}
\label{sec:method}

\subsection{Problem setting, operating point, and normalization}
\label{sec:method:problem}
\label{sec:method:model} % compatibility label (old name)

We consider a neutral-atom quantum processor based on optical tweezers.
A program is given as a logical quantum circuit $C$ composed of single-qubit gates and two-qubit $\mathrm{CZ}$ gates.
Our goal is to generate an \emph{executable schedule} $\mathcal{S}$ under geometry- and transport-related constraints, while optimizing a time--fidelity trade-off.

\textbf{Hardware abstraction.}
We assume a two-dimensional set of fixed trap sites.
Traps are divided into (i) \emph{home} traps (SLM traps) that initially host one atom per logical qubit, and (ii) \emph{hub} traps (AOD/auxiliary traps) that are initially empty and can temporarily store atoms during shuttling.
A $\mathrm{CZ}$ gate can be executed when the two participating atoms are within an effective Rydberg blockade radius.

\textbf{Operating point (physical parameters).}
To keep the compilation problem well-defined and to enable fair comparisons, we evaluate the compiler at a fixed device operating point:
the physical blockade radius is set to $r_b^{(\mathrm{phys})}=6~\mu\mathrm{m}$ (e.g., $^{87}$Rb with $n\approx 60$),
and the minimum safe separation between traps/atoms is set to $d_{\min}^{(\mathrm{phys})}=2~\mu\mathrm{m}$.
Other gate and transport timing/fidelity parameters are fixed and summarized in Table~\ref{tab:hwparams} (Sec.~\ref{sec:evaluation:hw}).

\textbf{Normalized coordinates and unit conversion.}
Our layout and hub placement are performed in a normalized coordinate system, where all trap coordinates lie in $[0,1]^2$.
Let $r_b$ denote the blockade radius in the normalized coordinates (this is \texttt{b\_rad} in our implementation).
We do \emph{not} assume that $r_b$ is a physical radius; instead, we map normalized distances to physical distances by the scale factor
\begin{equation}
s \;=\; \frac{r_b^{(\mathrm{phys})}}{r_b}\quad [\mu\mathrm{m}/\text{(normalized distance)}].
\label{eq:scale}
\end{equation}
Thus, any normalized distance $d$ corresponds to the physical distance $d^{(\mathrm{phys})}=s\,d$.
With this convention, the normalized minimum separation is $d_{\min}=d_{\min}^{(\mathrm{phys})}/s=(r_b/3)$, which matches the implementation choice \texttt{shuttle\_min\_sep = b\_rad/3}.

\begin{figure}[t]
    \centering
    \begin{minipage}[b]{0.32\linewidth}\centering
        \includegraphics[width=\linewidth]{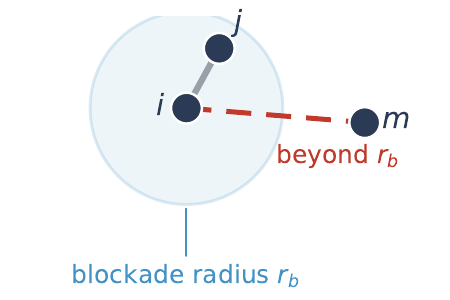}\\[2pt]
        {\small (a) Rydberg blockade}
    \end{minipage}\hfill
    \begin{minipage}[b]{0.32\linewidth}\centering
        \includegraphics[width=\linewidth]{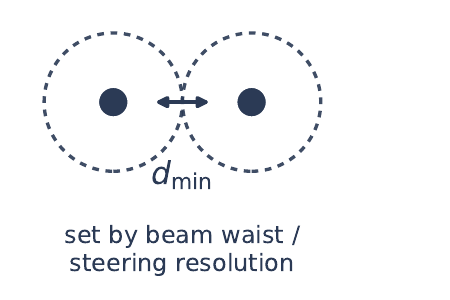}\\[2pt]
        {\small (b) Minimum separation}
    \end{minipage}\hfill
    \begin{minipage}[b]{0.32\linewidth}\centering
        \includegraphics[width=\linewidth]{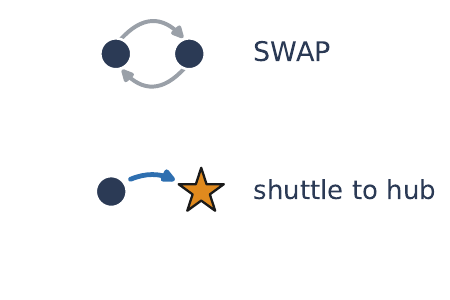}\\[2pt]
        {\small (c) Transport primitives}
    \end{minipage}
    \caption{Constraint model captured by our compiler: (a)~Rydberg blockade restricts feasible $\mathrm{CZ}$ interactions (within $r_b$) and parallel two-qubit operations, (b)~minimum separation $d_{\min}$ for safe addressing and motion, and (c)~transport primitives (SWAP and hub-mediated shuttling with optional eviction).}
    \label{fig:constraints}
\end{figure}

Fig.~\ref{fig:constraints} summarizes the constraint model used throughout the compiler.

\subsection{Overview: two-step compilation pipeline}
\label{sec:method:overview}

\begin{figure}[t]
    \centering
\includegraphics[width=\linewidth]{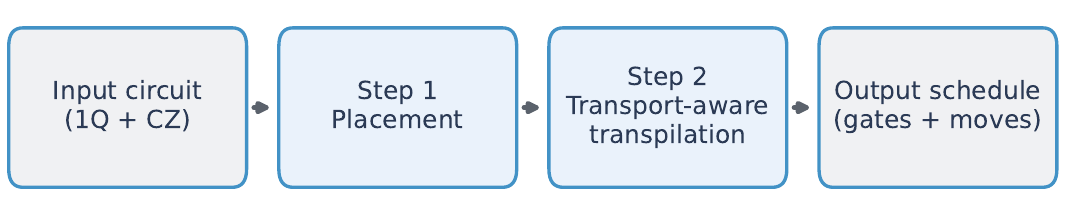}
    \caption{Two-step pipeline used in this work. Step~1 produces a circuit-aware placement and a normalized blockade radius. Step~2 adds hub traps and generates a schedule by selecting SWAP vs.\ shuttling/eviction on demand.}
    \label{fig:pipeline}
\end{figure}

Our compiler is organized as a \emph{two-step} pipeline (Fig.~\ref{fig:pipeline}):
\begin{enumerate}
    \item \textbf{Placement optimization (Step~1).} We optimize the 2D placement of logical qubits in normalized coordinates to reflect the $\mathrm{CZ}$ interaction structure of the input circuit, and determine a normalized blockade radius $r_b$ from the optimized layout.
    \item \textbf{Transport-aware transpilation (Step~2).} We insert hub traps and generate an executable schedule by dynamically choosing between SWAP-based routing and atom shuttling (with optional eviction) when a front-layer $\mathrm{CZ}$ is not directly executable.
\end{enumerate}

\begin{figure}[t]
  \centering
  \includegraphics[width=\columnwidth]{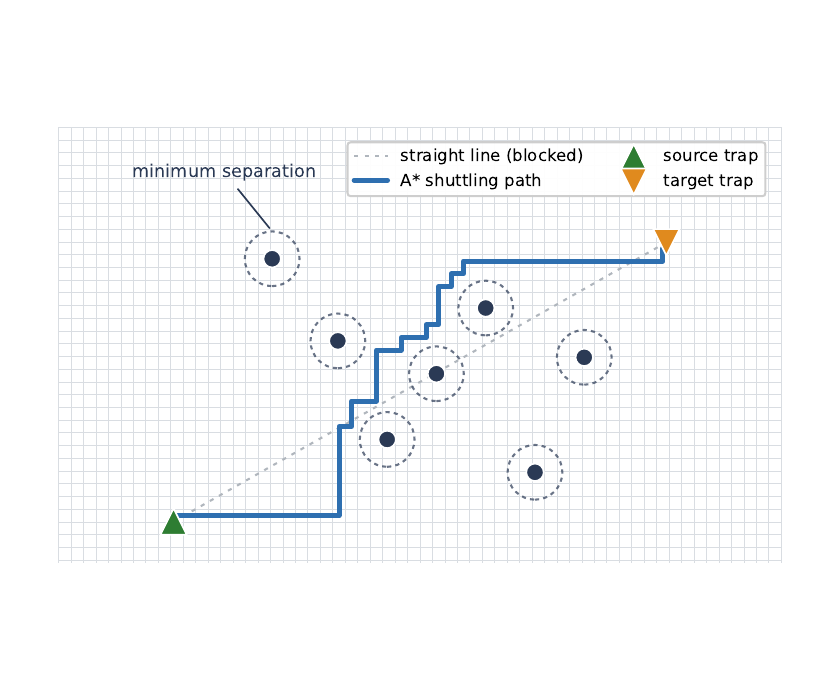}
  \caption{Grid-based A* precomputation for collision-avoiding shuttling distances.
  Exclusion zones around traps encode the minimum-separation constraint; pairwise distances are cached and reused during transpilation.}
  \label{fig:astar}
\end{figure}

Algorithm~\ref{alg:overall} and Fig.~\ref{fig:astar} summarize the end-to-end flow.

\begin{algorithm}[t]
\caption{Two-step neutral-atom compilation (implementation-aligned)}
\label{alg:overall}
\textbf{Input:} circuit $C$ (1Q + $\mathrm{CZ}$), number of hubs $N_{\mathrm{hub}}$, physical parameters (Table~\ref{tab:hwparams})\\
\textbf{Output:} schedule $\mathcal{S}$, normalized placement $X$, hub set $H$, normalized radius $r_b$
\begin{algorithmic}[1]
\STATE $(X,r_b)\leftarrow$ \textsc{PlacementOptimize}$(C;\ \texttt{maxiter}=10000,\ \texttt{seed}=0)$  \hfill \emph{// Step 1}
\STATE $H\leftarrow$ \textsc{HubPlace}$(C,X,r_b;\ N_{\mathrm{hub}},\ d_{\min}=r_b/3)$  \hfill \emph{// Step 2a}
\STATE Precompute collision-avoiding shuttling distances between all traps in $(X\cup H)$ using grid-based A* and memoize
\STATE $\mathcal{S}\leftarrow$ \textsc{TransportAwareTranspile}$(C,X,H,r_b)$ \hfill \emph{// Step 2b}
\STATE \textbf{return} $(\mathcal{S},X,H,r_b)$
\end{algorithmic}
\end{algorithm}

\subsection{Step 1: placement optimization and normalized blockade radius}
\label{sec:method:layout}
\label{sec:method:placement} % compatibility label (old name)

\textbf{Interaction graph extraction.}
We extract all $\mathrm{CZ}$ gates from the circuit and count the number of occurrences for each qubit pair.
This yields a weighted interaction graph $G=(V,E,w)$ where $w_{ij}$ is the $\mathrm{CZ}$ count between logical qubits $i$ and $j$.

\textbf{Placement objective (Graphine-style correlation).}
We assign each qubit $i\in V$ a 2D coordinate $\mathbf{x}_i\in[0,1]^2$.
The placement is optimized so that qubit pairs with larger $w_{ij}$ become geometrically closer.
Concretely, we minimize a normalized Pearson correlation between the weight vector $(w_{ij})$ and the distance vector $(\|\mathbf{x}_i-\mathbf{x}_j\|_2)$ over all pairs.
The resulting non-convex optimization is solved by a stochastic global optimizer (dual annealing) with \texttt{maxiter=10000} and \texttt{seed=0} (fixed for reproducibility).

\textbf{Selecting the normalized radius $r_b$.}
Given an optimized placement, we determine a normalized blockade radius $r_b$ by scanning candidate radii from the set of pairwise distances.
For each candidate radius $r$, we build a graph $G_r$ with edges for pairs within $r$.
We choose the smallest $r$ that satisfies:
(i) $G_r$ is connected, and
(ii) the hop-diameter of $G_r$ is at most $\sqrt{|V|}$.
If no candidate satisfies both conditions, we fall back to a minimum-spanning-tree based radius (maximum edge weight in the MST).

\textbf{Remark (physical interpretation).}
The value $r_b$ is a \emph{normalized} radius used internally for feasibility checks and for distance-to-time conversion via Eq.~\eqref{eq:scale}.
All experiments share the same physical blockade radius $r_b^{(\mathrm{phys})}=6~\mu\mathrm{m}$; the normalization simply determines the scale factor for the layout.
\subsection{Step 2a: hub trap placement from \texorpdfstring{$\mathrm{CZ}$}{CZ} geometry}
%\subsection{Step 2a: hub trap placement from $\mathrm{CZ}$ geometry}
\label{sec:method:hubs}

\begin{figure}[t]
  \centering
  \includegraphics[width=\columnwidth]{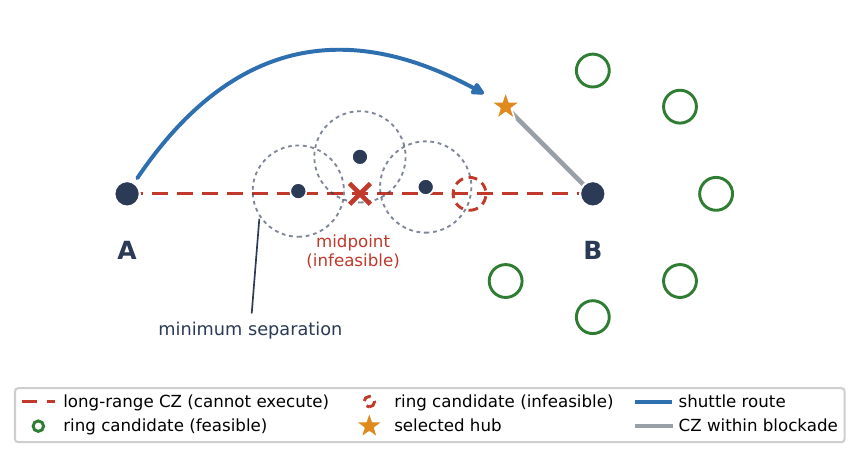}
  \caption{Illustration of the hub placement heuristic (strategic2).
  Candidate sites are generated from CZ geometry (midpoints of long-range CZ pairs) and filtered by the minimum-separation constraint.
  Feasible candidates are greedily selected by the endpoint-proximity score weighted by CZ counts.}
  \label{fig:hubplacement}
\end{figure}

We place $N_{\mathrm{hub}}$ empty hub traps to reduce transport overhead for long-range interactions.
Our hub placement heuristic (used in the reported results) is described in Algorithm~\ref{alg:hubplace} and visualized in Fig.~\ref{fig:hubplacement}.

% --- Fig: hub placement heuristic (insert right after Algorithm 2) ---

\textbf{Candidate generation.}
From the placement $X$, we compute the distance for each $\mathrm{CZ}$ pair.
We define \emph{long-range} $\mathrm{CZ}$ pairs as those whose distance exceeds $1.1\,r_b$ in normalized units.
We then use the midpoints of long-range $\mathrm{CZ}$ pairs as hub candidates.
If no long-range $\mathrm{CZ}$ exists, we use midpoints of all $\mathrm{CZ}$ pairs.

\textbf{Feasibility filtering and greedy selection.}
A candidate is feasible if it is at least $d_{\min}=r_b/3$ away from every SLM atom and from every previously selected hub.
Each feasible candidate is assigned a score that favors candidates close to the endpoints of frequently used long-range $\mathrm{CZ}$ interactions:
\begin{equation}
\mathrm{score}(c) = \sum_{(i,j)\in E_{\mathrm{long}}} w_{ij}\!\left(\frac{1}{1+\|\mathbf{x}_i-c\|_2}+\frac{1}{1+\|\mathbf{x}_j-c\|_2}\right).
\label{eq:hubscore}
\end{equation}
We greedily pick the highest-score feasible candidate until $N_{\mathrm{hub}}$ hubs are selected or no feasible candidates remain.

\begin{algorithm}[t]
\caption{\textsc{HubPlace} heuristic used in this work (\texttt{strategic2})}
\label{alg:hubplace}
\textbf{Input:} circuit $C$, normalized placement $X=\{\mathbf{x}_i\}$, normalized radius $r_b$, hubs $N_{\mathrm{hub}}$, minimum separation $d_{\min}=r_b/3$\\
\textbf{Output:} hub positions $H$ (initially empty traps)
\begin{algorithmic}[1]
\STATE Count $\mathrm{CZ}$ occurrences per pair $(i,j)$ to obtain weights $w_{ij}$
\STATE $E_{\mathrm{long}} \leftarrow \{(i,j)\mid \|\mathbf{x}_i-\mathbf{x}_j\|_2 > 1.1\,r_b\}$; if empty, set $E_{\mathrm{long}}\leftarrow E$
\STATE Candidate set $\mathcal{C}\leftarrow \{(\mathbf{x}_i+\mathbf{x}_j)/2 \mid (i,j)\in E_{\mathrm{long}}\}$
\STATE $H\leftarrow \emptyset$
\WHILE{$|H|<N_{\mathrm{hub}}$ and $\mathcal{C}\neq \emptyset$}
    \STATE Remove $c\in\mathcal{C}$ that violates distance constraints to atoms or hubs ($<d_{\min}$)
    \STATE Compute $\mathrm{score}(c)$ by Eq.~\eqref{eq:hubscore}
    \STATE Select $c^\star \leftarrow \arg\max_{c\in\mathcal{C}} \mathrm{score}(c)$ and add it to $H$
    \STATE Remove $c^\star$ from $\mathcal{C}$
\ENDWHILE
\STATE \textbf{return} $H$
\end{algorithmic}
\end{algorithm}

\subsection{Endpoint-ring extension to hub candidate generation}
\label{sec:method:endpoint-ring}

The midpoint candidate set $\mathcal{C}$ in Algorithm~\ref{alg:hubplace} can be exhausted before $N_{\mathrm{hub}}$ feasible hubs are found.
This happens when the midpoint of a long-range $\mathrm{CZ}$ pair sits closer than $d_{\min}$ to an existing atom or hub, so the only natural candidate for that pair is rejected.
When this occurs for a pair whose two endpoints are far apart, the scheduler is left with no transit waypoint and falls back to SWAP routing for that pair.

The endpoint-ring extension enlarges the candidate set without changing the hub budget $N_{\mathrm{hub}}$.
For each long-range $\mathrm{CZ}$ pair, in addition to the midpoint, we generate ring candidates placed around each endpoint at radius $0.9\,r_b$ in eight evenly spaced directions.
A ring candidate sits within blockade range of its endpoint and provides an alternative transit point when the midpoint is infeasible.
The greedy selection in Algorithm~\ref{alg:hubplace} is otherwise unchanged.
Ring candidates enter the same feasibility filter and the same score in Eq.~\eqref{eq:hubscore}, and the loop still stops at $N_{\mathrm{hub}}$ hubs.
In the ablation of Sec.~\ref{sec:evaluation}, the method labeled \texttt{Proposed} disables this extension and \texttt{Proposed+Ring} enables it.

\subsection{Step 2b: transport-aware transpilation (SWAP vs.\ shuttling/eviction)}
\label{sec:method:transport}
\label{sec:method:transpilation} % compatibility label (old name)

We generate an executable schedule by processing the circuit in a dependency-respecting order and inserting transport operations only when necessary.
We maintain (i) a mapping from atoms to occupied traps and (ii) a set of empty traps (hubs and any vacated home traps).

\textbf{Feasibility of a $\mathrm{CZ}$.}
A $\mathrm{CZ}$ between atoms $a$ and $b$ is executable if their current trap positions are within the normalized blockade radius $r_b$.
In addition, our scheduler enforces blockade-limited parallelism: two $\mathrm{CZ}$ operations cannot be placed in the same time layer if any cross-pair distance is within the effective blockade radius.

\textbf{Transport choices.}
When a front-layer $\mathrm{CZ}$ is not executable, we insert transport operations to bring the atoms within $r_b$.
We consider:
\begin{itemize}
    \item \textbf{SWAP routing (SLM-only).} We treat the current occupied SLM traps as a dynamic coupling graph with edges between pairs within $r_b$. We move quantum states by applying SWAPs along a shortest path.
    \item \textbf{Shuttling (with hubs).} We move an atom to an empty trap (hub or home) that lies within $r_b$ of the partner atom. Shuttling distances are evaluated by collision-avoiding A* path lengths precomputed on a grid.
    \item \textbf{Lazy eviction.} If a promising target trap is occupied by an \emph{idle} atom (not in the current ready set) that is away from its home trap, we recursively evict it back to its home to free the target, and then perform the shuttle.
\end{itemize}

\subsection{Cost model and decision rule}
\label{sec:method:cost}

To decide between SWAP and shuttling plans, we use a scalar score that balances execution time and operation fidelity.
For a candidate plan with total duration $t$ and per-operation fidelities $\{f_k\}$, we compute
\begin{equation}
S \;=\; \frac{t}{T_{\mathrm{eff}}}\;-\;\sum_k \log f_k,
\label{eq:score}
\end{equation}
and optionally apply multiplicative scaling factors $\alpha_g$ (for SWAP-based plans) and $\alpha_s$ (for shuttling-based plans) as $S/\alpha$.
In our current experiments, we set $\alpha_g=\alpha_s=1$.

In the implementation, a SWAP is modeled as a fixed decomposition cost of $\texttt{cz\_per\_swap}=3$ and $\texttt{oneq\_per\_swap}=4$ elementary gates.
A shuttling operation with normalized distance $d$ is modeled as
\begin{equation}
t_{\mathrm{sh}}(d)=2t_{\mathrm{act}}+\frac{s\,d}{v_{\mathrm{sh}}^{(\mathrm{phys})}},
\label{eq:shuttle_time}
\end{equation}
where $s$ is the scale factor in Eq.~\eqref{eq:scale} and $v_{\mathrm{sh}}^{(\mathrm{phys})}$ is the physical shuttling speed.
Algorithm~\ref{alg:decision} summarizes the decision procedure.

\begin{algorithm}[t]
\caption{\textsc{TransportDecision} for a blocked $\mathrm{CZ}$ (SWAP vs.\ shuttling/eviction)}
\label{alg:decision}
\textbf{Input:} atoms $(a,b)$ for the pending $\mathrm{CZ}$, current occupancy, ready-set atoms $A_{\mathrm{ready}}$\\
\textbf{Output:} transport plan $\mathcal{P}$ (sequence of SWAPs and/or shuttles)
\begin{algorithmic}[1]
\STATE Construct a SWAP candidate along a shortest path in the current SLM coupling graph; compute its score $S_{\mathrm{swap}}$
\STATE Initialize best shuttling score $S_{\mathrm{sh}}\leftarrow+\infty$
\FOR{direction in $\{a\rightarrow b,\ b\rightarrow a\}$}
    \STATE Enumerate target traps $t$ within distance $r_b$ of the stationary atom
    \FOR{each target $t$}
        \IF{$t$ is empty}
            \STATE Candidate plan: shuttle the moving atom to $t$
        \ELSIF{$t$ is occupied by an idle atom $c\notin A_{\mathrm{ready}}$ away from its home}
            \STATE Candidate plan: evict $c$ to its home (possibly recursively), then shuttle
        \ENDIF
        \STATE Evaluate the candidate score $S$ using Eq.~\eqref{eq:score} with shuttling distance from the A* cache
        \STATE Update $(S_{\mathrm{sh}},\mathcal{P}_{\mathrm{sh}})$ if improved
    \ENDFOR
\ENDFOR
\STATE \textbf{return} the plan with smaller score (tie: optionally prefer shuttling)
\end{algorithmic}
\end{algorithm}

\subsection{Serial and parallel shuttle models}
\label{sec:method:shuttle-models}

The decision rule above schedules shuttle operations one at a time, which we refer to as the serial shuttle model (M0).
M0 charges the full duration of Eq.~\eqref{eq:shuttle_time} to every shuttle and is the model used for the headline execution-time figures.

Reconfigurable neutral-atom hardware can move multiple atoms within a single AOD transfer, provided their source-destination segments do not intersect.
We capture this with a parallel shuttle model (M2) implemented as an offline post-processing pass over the M0 schedule.
M2 batches shuttle operations whose travel segments are mutually non-intersecting and that lie within a contiguous window of layers in which no other operation acts on the shuttled atoms.
Each batch is then charged a single transfer duration rather than the sum of its members.
M2 does not change gate counts, layer counts, or fidelity.
It only lowers the execution-time estimate, and we report it as a sensitivity in Sec.~\ref{sec:evaluation:sensitivities} rather than as a primary metric.

\subsection{Implementation note}
\label{sec:method:impl}

For reproducibility and for 1-to-1 correspondence with our prototype, we summarize the mapping between the above description and the implementation:
(i) Step~1 is implemented by \circuit{optimize\_atom\_positions\_da\_connected\_new},
(ii) Step~2a is implemented by \circuit{\_calculate\_hub\_positions\_strategic2},
and (iii) Step~2b (layering, blockade checks, and transport insertion) is implemented in the \circuit{RydbergBlockadeEstimator} class.

\section{Evaluation}
\label{sec:evaluation}

\subsection{Experimental setup and benchmarks}
\label{sec:evaluation:setup}

\textbf{Prototype implementation.}
We evaluate a Python prototype built on top of Qiskit~\cite{javadiabhari2024qiskit}.
Unless otherwise stated, the placement optimizer uses \texttt{maxiter=10000} and \texttt{seed=0} for deterministic reproduction.
All experiments are run with Python~3.12.3, Qiskit~1.1.1, SciPy, and NetworkX~$\ge$~3.3.

\textbf{Benchmarks.}
The seventeen circuits in Table~\ref{tab:benchmarks} are taken from the benchmark suite distributed with Parallax~\cite{ludmir2024parallax}, already expressed in the single-qubit and $\mathrm{CZ}$ gate set that our compiler consumes.
Reusing an established neutral-atom compilation suite, rather than a hand-picked set, keeps the benchmark selection independent of the method under study.
The suite spans two axes that drive the compilation problem we consider.
The first is scale, with qubit counts from 9 to 39 across the small and medium ranges that current monolithic devices target.
The second is the structure of the interaction graph, ranging from circuits whose $\mathrm{CZ}$ graph is local enough to be absorbed by placement alone to circuits whose long-range $\mathrm{CZ}$ demand forces routing, which is the distinction Sec.~\ref{sec:evaluation:main} uses to separate the two regimes.
We use seventeen of the eighteen circuits in the suite and omit the 128-qubit transverse-field Ising instance, which lies above the qubit-count range we study.
For each circuit we report the total number of gates and the number of $\mathrm{CZ}$ gates, as these directly drive routing and transport overhead.

\begin{table}[t]
\centering
\caption{Benchmark circuits used in Sec.~\ref{sec:evaluation}. Gate counts are extracted from the OpenQASM source.}
\label{tab:benchmarks}
\footnotesize
\setlength{\tabcolsep}{4pt}
\renewcommand{\arraystretch}{1.06}
\begin{tabularx}{\linewidth}{>{\raggedright\arraybackslash}X r r r}
\toprule
Circuit & \#Qubits & \#CZ & \#Gates \\
\midrule
\circuit{adder_9} & 9 & 153 & 415 \\
\circuit{advantage_9} & 9 & 32 & 86 \\
\circuit{hlf_10} & 10 & 54 & 141 \\
\circuit{multiplier_10} & 10 & 293 & 827 \\
\circuit{qaoa_10} & 10 & 162 & 457 \\
\circuit{qft_10} & 10 & 444 & 1249 \\
\circuit{seca_n11} & 11 & 80 & 194 \\
\circuit{sat_11} & 11 & 252 & 609 \\
\circuit{gcm_h6_13} & 13 & 528 & 1372 \\
\circuit{heisenberg_16} & 16 & 3081 & 8681 \\
\circuit{qec9xz_n17} & 17 & 32 & 99 \\
\circuit{sqrt_18} & 18 & 898 & 2102 \\
\circuit{knn_n25} & 25 & 84 & 230 \\
\circuit{wstate_27} & 27 & 52 & 157 \\
\circuit{vqe_uccsd_n28} & 28 & 194596 & 495955 \\
\circuit{qv_32} & 32 & 1488 & 4496 \\
\circuit{qugan_n39} & 39 & 205 & 579 \\
\bottomrule
\end{tabularx}
\end{table}

\begin{figure}[t]
  \centering
  \includegraphics[width=\columnwidth]{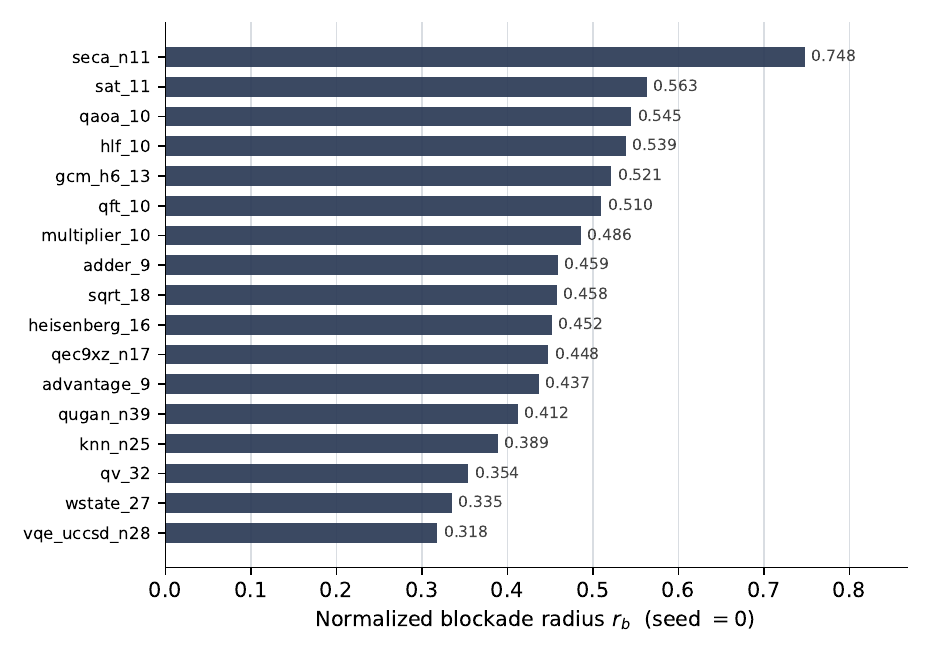}
  \caption{Distribution of the normalized blockade radius \(r_b\) returned by the placement optimizer (seed=0).
  The physical operating point fixes \(r_b^{(\mathrm{phys})}=6\,\mu\mathrm{m}\); \(r_b\) determines the distance scaling factor.}
  \label{fig:rbdist}
\end{figure}

Fig.~\ref{fig:rbdist} visualizes how the normalized blockade radius varies across these circuits.

\subsection{Operating point and hardware parameters}
\label{sec:evaluation:hw}

Table~\ref{tab:hwparams} summarizes the physical parameters used in the execution-time and fidelity estimates.
As discussed in Sec.~\ref{sec:method:problem}, compilation is performed in normalized coordinates with a normalized blockade radius $r_b$ returned by Step~1.
For evaluation, normalized distances $d$ are converted to physical distances by Eq.~\eqref{eq:scale}, i.e., $d^{(\mathrm{phys})}=s\,d$ with $s=r_b^{(\mathrm{phys})}/r_b$.
With this convention, the minimum safe separation $d_{\min}^{(\mathrm{phys})}=2~\mu\mathrm{m}$ corresponds to the normalized constraint $d_{\min}=r_b/3$ used by the transpiler.

\begin{table}[t]
\centering
\caption{Fixed operating point used in evaluation. Times are in $\mu$s and distances are in $\mu$m.}
\label{tab:hwparams}
\footnotesize
\setlength{\tabcolsep}{4pt}
\renewcommand{\arraystretch}{1.06}
\begin{tabular}{l@{\hspace{0.8em}}l}
\toprule
Parameter & Value \\
\midrule
Physical blockade radius $r_b^{(\mathrm{phys})}$ & $6~\mu\mathrm{m}$ \\
Minimum separation $d_{\min}^{(\mathrm{phys})}$ & $2~\mu\mathrm{m}$ \\
1Q gate duration $t_{1Q}$ & $2.0~\mu\mathrm{s}$ \\
$\mathrm{CZ}$ duration $t_{CZ}$ & $0.8~\mu\mathrm{s}$ \\
Trap (de)activation time $t_{\mathrm{act}}$ & $100~\mu\mathrm{s}$ \\
Shuttling speed $v_{\mathrm{sh}}^{(\mathrm{phys})}$ & $0.55~\mu\mathrm{m}/\mu\mathrm{s}$ \\
1Q fidelity $F_{1Q}$ & $0.999$ \\
$\mathrm{CZ}$ fidelity $F_{CZ}$ & $0.995$ \\
Relaxation time $T_1$ & $1.0\times 10^8~\mu\mathrm{s}$ \\
Dephasing time $T_2$ & $1.5\times 10^6~\mu\mathrm{s}$ \\
Score scaling $\alpha_g,\alpha_s$ & $1.0,1.0$ \\
\bottomrule
\end{tabular}
\end{table}

\subsection{Metrics}
\label{sec:evaluation:metrics}

All metrics in this section are analytic estimates computed from the compiled schedule rather than hardware measurements or full state-vector simulation.
The execution time and the fidelity proxy are deterministic functions of the layered schedule and the operating point in Tab.~\ref{tab:hwparams}.

\textbf{Estimated execution time.}
Given a compiled schedule represented as a sequence of time layers, we estimate the total execution time as the sum of the maximum operation duration in each layer.
Single-qubit and $\mathrm{CZ}$ gates use the fixed durations in Table~\ref{tab:hwparams}.
A SWAP is modeled using a fixed decomposition cost (three $\mathrm{CZ}$ and four single-qubit gates per SWAP, consistent with the cost model in Sec.~\ref{sec:method:cost}).
A shuttling operation with normalized travel distance $d$ is charged the duration $t_{\mathrm{sh}}(d)$ defined in Eq.~\eqref{eq:shuttle_time}.

\textbf{Estimated circuit fidelity (proxy).}
We report a multiplicative proxy fidelity obtained by combining gate fidelities and a decoherence factor per layer.
For each layer with maximum duration $t_{\max}$, we multiply by $\exp(-t_{\max}/T_{\mathrm{eff}})$ where $T_{\mathrm{eff}}=(T_1T_2)/(T_1+T_2)$,
and we multiply by $F_{1Q}$ or $F_{CZ}$ for each 1Q or $\mathrm{CZ}$ gate in the layer, respectively.
For each shuttle operation in the layer we multiply by a per-shuttle factor $F_{\mathrm{sh}}$ whose default value is $F_{\mathrm{sh}}=1$.
The $T_2$ dephasing component of the transport cost is already captured by the layer-time decoherence factor; $F_{\mathrm{sh}}$ is the operation-induced loss (heating, motional excitation, transfer error) that this factor isolates.
Sec.~\ref{sec:evaluation:sensitivities} reports a sweep of $F_{\mathrm{sh}}$ over $\{1,\,0.999,\,0.99\}$.

\subsection{Method matrix}
\label{sec:evaluation:method-matrix}

We compare seven methods: the full pipeline (\circuit{Proposed+Ring}), three ablations of it (\circuit{Proposed}, \circuit{No-Eviction}, \circuit{No-Hub}), the closest published baseline (\circuit{Graphine(SWAP-only)}), and two Qiskit-default reference points (TLS and PTR).
The full configuration is given in Table~\ref{tab:method_matrix}.
All proposed methods use fully automatic hub placement; manual hub overrides used during prototyping are excluded from all results below.

\begin{table}[t]
\centering
\caption{Configurations of our pipeline and its ablations. \circuit{Proposed+Ring} is the full method; the remaining rows disable, in turn, the endpoint-ring extension (\circuit{Proposed}), eviction shuttling (\circuit{No-Eviction}), and hub traps entirely (\circuit{No-Hub}). We additionally compare against three hub-free references, all with $N_{\mathrm{hub}}=0$ and SWAP-only routing: \circuit{Graphine(SWAP-only)}, our reading of~\cite{patel2023graphine} and parameter-identical to \circuit{No-Hub}; \circuit{Trivial-Layout-SWAP} (TLS), the Qiskit default on a regular grid; and \circuit{Placement-Trivial-Routing} (PTR), Qiskit SABRE routing on our Step~1 placement.}
\label{tab:method_matrix}
\footnotesize
\setlength{\tabcolsep}{5pt}
\renewcommand{\arraystretch}{1.2}
\begin{tabular}{@{}lccc@{}}
\toprule
Method & $N_{\mathrm{hub}}$ & Eviction & Endpoint ring \\
\midrule
\circuit{Proposed+Ring} & 8 & yes & yes \\
\circuit{Proposed}      & 8 & yes & no  \\
\circuit{No-Eviction}   & 8 & no  & no  \\
\circuit{No-Hub}        & 0 & no  & no  \\
\bottomrule
\end{tabular}
\end{table}

\textbf{Headline method.}
The headline method is \circuit{Proposed+Ring}: the full pipeline with hub traps, eviction shuttling, and the endpoint-ring extension to hub candidate generation (Sec.~\ref{sec:method:endpoint-ring}).

\textbf{Ablations.}
\circuit{Proposed} disables the endpoint-ring extension, isolating its contribution.
\circuit{No-Eviction} keeps hub traps but disables eviction shuttling.
\circuit{No-Hub} disables both, leaving only SWAP-based routing on top of our Step~1 placement.

\textbf{Literature baseline.}
\circuit{Graphine(SWAP-only)} is parameter-identical to \circuit{No-Hub} but flagged separately as our reading of the closest published baseline~\cite{patel2023graphine}.

\textbf{Qiskit references.}
\circuit{Trivial-Layout-SWAP} is the Qiskit default end-to-end pipeline applied to a regular grid layout.
\circuit{Placement-Trivial-Routing} reuses our Step~1 placement and applies Qiskit's default SWAP-based routing, the SABRE heuristic~\cite{li2019sabre} in its LightSABRE implementation~\cite{zou2024lightsabre}, on top, isolating the routing-side improvement at fixed placement.
\subsection{Main results}
\label{sec:evaluation:main}

\begin{figure}[t]
  \centering
  \includegraphics[width=\linewidth]{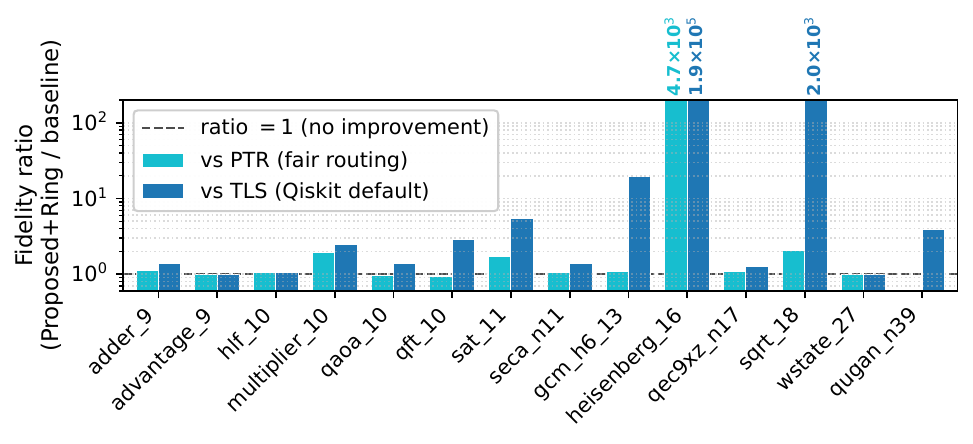}
\caption{Per-benchmark fidelity ratio of \circuit{Proposed+Ring} against two baselines, at $F_{\mathrm{sh}}=1$.
  The dashed line in the legend marks ratio $1$, above which \circuit{Proposed+Ring} improves on the baseline and below which the baseline is better.
  Bars are capped at $200$; the three ratios that exceed the cap (\circuit{heisenberg_16} against PTR and TLS, and \circuit{sqrt_18} against TLS) are printed above their bars.
  The large gains are concentrated on routing-dominated circuits; for most benchmarks the ratio against PTR lies between $0.7\!\times$ and $2\!\times$.
  PTR fails to compile \circuit{qugan_n39} on Qiskit's coupling-map check, so only the TLS ratio is shown for that circuit, and \circuit{vqe_uccsd_n28} is excluded as a numerical scope-out case (Sec.~\ref{sec:evaluation:main}).}
  \label{fig:eval:fidelity_improvement}
\end{figure}

\textbf{Method comparison overview.}
Figure~\ref{fig:eval:fidelity_improvement} shows the per-benchmark fidelity ratio of \circuit{Proposed+Ring} against two reference baselines.
\circuit{Placement-Trivial-Routing} (PTR) reuses our Step~1 placement and applies Qiskit's default SWAP-based routing on top.
\circuit{Trivial-Layout-SWAP} (TLS) is the Qiskit default end-to-end pipeline on a regular grid layout.
The PTR comparison isolates the routing improvement at fixed placement.
The TLS comparison gives the end-to-end improvement that an unmodified Qiskit user would see.

Fifteen of the seventeen benchmarks complete under \circuit{Proposed+Ring} within the 15-minute compile budget.
On these fifteen, the distribution of fidelity ratios is bimodal.
Two benchmarks reach gains of three orders of magnitude or more.
\circuit{heisenberg_16} reaches $4.73\!\times\!10^{3}$ against PTR and $1.90\!\times\!10^{5}$ against TLS.
\circuit{sqrt_18} reaches $2.04\!\times$ against PTR and $1.98\!\times\!10^{3}$ against TLS.
Because \circuit{heisenberg_16} is by far the deepest completed circuit, its absolute proxy fidelities are small for all methods (\circuit{Proposed+Ring} $7.1\!\times\!10^{-10}$, PTR $1.5\!\times\!10^{-13}$, TLS $3.7\!\times\!10^{-15}$), so this ratio measures the reduction in routing overhead rather than a usable fidelity.
A third benchmark, \circuit{qugan_n39}, completes only under the proposed methods.
PTR aborts on \circuit{qugan_n39} at Qiskit's coupling-map check.\footnote{Qiskit raises ``A connected component of the DAGCircuit is too large for any of the connected components in the coupling map.'' This traces to the placement producing a contact graph sparser than what the circuit DAG requires of a SWAP-only router.}
Only the TLS ratio is therefore shown for \circuit{qugan_n39} in Fig.~\ref{fig:eval:fidelity_improvement}.
The remaining benchmarks fall between $0.7\!\times$ and $2\!\times$ against PTR.
We explain this clustering below.

\begin{figure}[t]
  \centering
  \includegraphics[width=\linewidth]{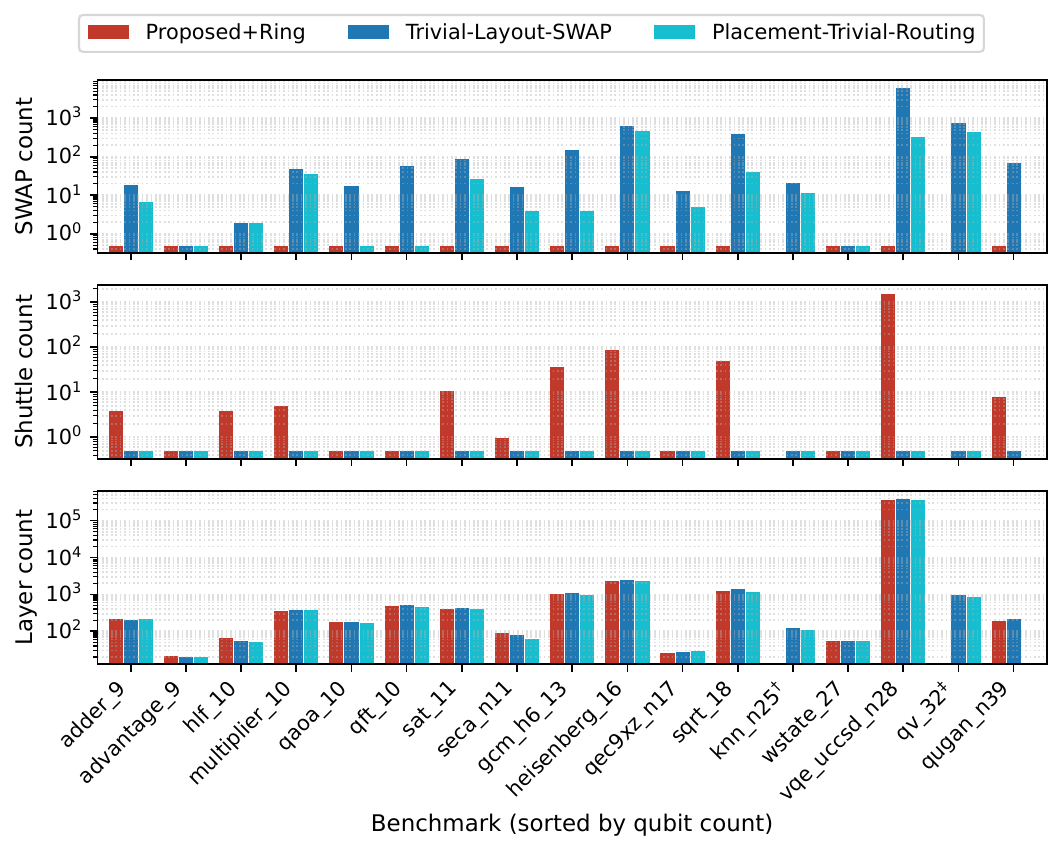}
  \caption{Per-benchmark resource breakdown for three methods (log scale): \circuit{Proposed+Ring}, \circuit{Trivial-Layout-SWAP} (TLS), and \circuit{Placement-Trivial-Routing} (PTR). Top: SWAP count. Middle: shuttle count. Bottom: layer count. Zero values are floored to $0.5$ for display.
  \circuit{Proposed+Ring} issues zero SWAP gates on every completed circuit and routes through shuttling instead, whereas the two SWAP-only baselines issue no shuttles.
  A dagger ($\dagger$) marks \circuit{knn_n25}, on which \circuit{Proposed+Ring} times out, and a double dagger ($\ddagger$) marks \circuit{qv_32}, on which it returns a no-valid-transport error; both are analyzed in Sec.~\ref{sec:evaluation:failure-modes}.}
  \label{fig:eval:breakdown}
\end{figure}

\begin{figure}[t]
  \centering
  \includegraphics[width=\columnwidth]{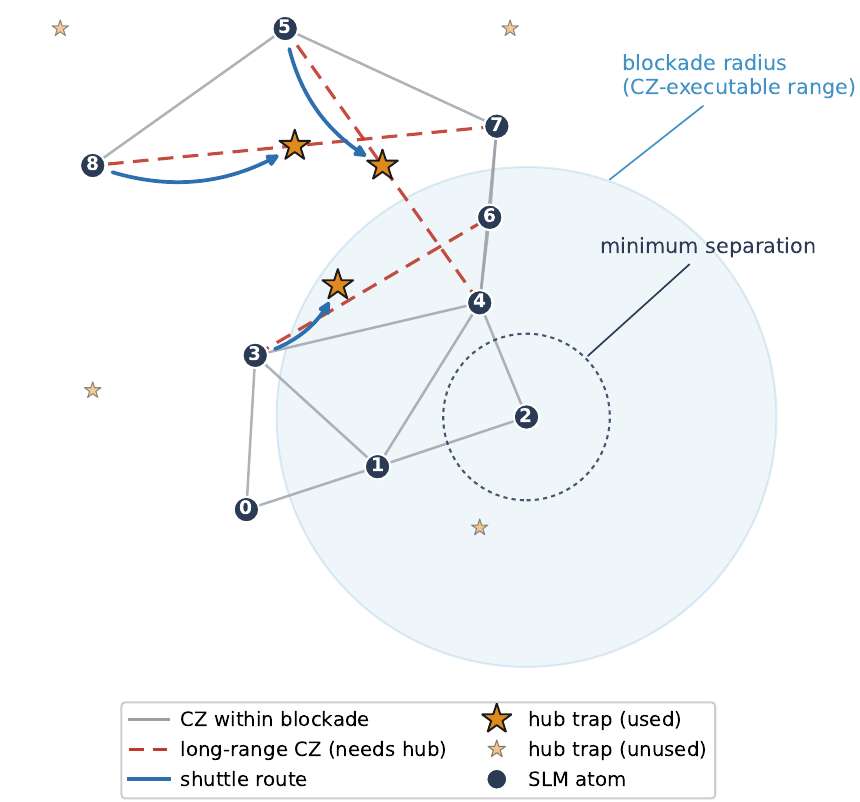}
\caption{Application-specific layout and hub-assisted routing for the
\texttt{adder\_9} benchmark compiled with the proposed method. Solid lines
mark CZ interactions whose qubits are placed within the blockade radius and
therefore execute directly, while dashed lines mark long-range CZ interactions
whose qubits lie beyond it and cannot be driven without relocating an atom.
The shaded disk shows the blockade radius for one representative atom and the
inner dotted circle shows the minimum trap separation. Stars mark hub traps.
The three hubs used in this instance each receive an atom shuttled from its
home trap (arrows) to within the blockade radius of its partner, where the CZ
then executes locally.}
  \label{fig:case_study}
\end{figure}

\textbf{Source of the gain.}
Figure~\ref{fig:eval:breakdown} shows where the gain comes from.
On every benchmark for which \circuit{Proposed+Ring} completes, it issues zero SWAP gates and routes through hub-mediated and eviction shuttling.
For \circuit{heisenberg_16}, the 669 SWAPs issued by TLS are replaced by 91 shuttles.
For \circuit{sqrt_18}, the 402 SWAPs are replaced by 50 shuttles.
Each SWAP decomposes into three $\mathrm{CZ}$ gates at $F_{CZ}=0.995$ (Tab.~\ref{tab:hwparams}), so on circuits with substantial routing the fidelity effect of removing SWAPs is large.
Figure~\ref{fig:case_study} shows the compiled layout for \texttt{adder\_9}.
The placement stage, which builds on the force-directed embedding of
Graphine~\cite{patel2023graphine}, draws frequently interacting qubits together, so that
twelve of the fifteen distinct CZ interactions fall within the blockade radius
and execute without any atom movement. The remaining three interactions are
long-range. Their qubits are separated by more than the blockade radius, and
since the Rydberg interaction that drives a CZ falls off sharply beyond this
range, the gate cannot be realized by scheduling alone. This is different from
the blockade interference that limits parallelism, which is resolved by
serializing gates in time; a long-range CZ instead requires one of its qubits
to be physically relocated. For each such interaction the compiler weighs the
cost of a SWAP chain through neighboring traps against shuttling one qubit to
an auxiliary hub trap, and takes the cheaper option. Here all three long-range
CZ are handled by shuttling, with one qubit moved from its home trap to a hub
that lies within the blockade radius of its partner, where the CZ then executes
directly. Seven hub traps are placed from a budget of eight, of which three are
used in this instance while the rest remain available for other interactions.

\textbf{Routing-required and routing-free regimes.}
The benchmark suite separates into four regimes by the structure of the $\mathrm{CZ}$ interaction graph relative to our placement.
\emph{Regime~A (routing-dominated, large benefit)}: \circuit{heisenberg_16} and \circuit{sqrt_18}, the two circuits with by far the largest routing load.
Both have $\mathrm{CZ}$ graphs whose diameter scales with the qubit count and a high fraction of qubit pairs that cannot be resolved within blockade.
Among the benchmarks that complete under \circuit{Proposed+Ring} and enter Fig.~\ref{fig:eval:fidelity_improvement}, these two carry by far the largest routing load, with \circuit{heisenberg_16} at 3081 $\mathrm{CZ}$ gates and \circuit{sqrt_18} at 898 against a next-largest count of 528 (Tab.~\ref{tab:benchmarks}).
Their position at the top of the figure therefore tracks routing demand rather than a favorable choice of instance.
The endpoint-ring extension (Sec.~\ref{sec:method:endpoint-ring}) is required here.
Without it, the strategic2 hub allocator cannot place a feasible midpoint candidate under the minimum-separation constraint, and the proposed method falls back to the No-Hub regime.
\emph{Regime~B (modest routing benefit, up to ${\sim}2\!\times$ vs.\ PTR)}: \circuit{sat_11} ($1.70\!\times$ vs.\ PTR), \circuit{multiplier_10} ($1.94\!\times$), and \circuit{vqe_uccsd_n28} ($1.92\!\times$, see ``Excluded benchmark'' below). \circuit{gcm_h6_13} ($1.07\!\times$) is a boundary case: its benefit is marginal and disappears under pessimistic shuttle fidelity (Sec.~\ref{sec:evaluation:sensitivities}).
\emph{Regime~C (no measurable benefit, near $1\!\times$)}: \circuit{adder_9}, \circuit{hlf_10}, \circuit{seca_n11}, \circuit{qec9xz_n17}.
The $\mathrm{CZ}$ graphs are local enough that our placement absorbs almost all interactions within blockade, leaving zero to four shuttles per circuit.
\emph{Regime~D (no routing required)}: \circuit{advantage_9}, \circuit{wstate_27}, \circuit{qaoa_10}, \circuit{qft_10}.
Our placement maps the entire circuit within blockade, and all methods produce numerically identical results.
\circuit{qugan_n39} completes under the proposed methods but provides no PTR ratio (PTR aborts at Qiskit's coupling-map check), so it is reported via its TLS ratio in Fig.~\ref{fig:eval:fidelity_improvement} and is not assigned a PTR-ratio-based regime.

The bimodality of Fig.~\ref{fig:eval:fidelity_improvement} reflects this structural property of the suite.
The proposed compiler intervenes meaningfully on circuits whose interaction graph has long-range edges that cannot be resolved by placement alone.

\begin{table}[t]
\centering
\caption{Hub traps as an enabling capability. All ten circuits listed already exceed the 15-minute per-method budget under the SWAP-only \circuit{No-Hub} ablation (the selection criterion); we report the \circuit{No-Hub} outcome under an extended two-hour budget and the \circuit{Proposed+Ring} outcome under the 15-minute budget. \circuit{No-Hub} returns no schedule at two hours on every circuit, whereas \circuit{Proposed+Ring} compiles seven of them with zero SWAP gates. Two circuits (\circuit{knn_n25}, \circuit{qv_32}) and the numerical scope-out case (\circuit{vqe_uccsd_n28}) are treated separately in Sec.~\ref{sec:evaluation:failure-modes}.}
\label{tab:enabling}
\footnotesize
\setlength{\tabcolsep}{4pt}
\renewcommand{\arraystretch}{1.15}
\begin{tabular}{@{}lcl@{}}
\toprule
Circuit & No-Hub (2\,h) & Proposed+Ring \\
\midrule
\circuit{adder_9}        & timeout & 0 SWAP, 4 shuttles \\
\circuit{multiplier_10}  & timeout & 0 SWAP, 5 shuttles \\
\circuit{sat_11}         & timeout & 0 SWAP, 11 shuttles \\
\circuit{gcm_h6_13}      & timeout & 0 SWAP, 38 shuttles \\
\circuit{heisenberg_16}  & timeout & 0 SWAP, 91 shuttles \\
\circuit{sqrt_18}        & timeout & 0 SWAP, 50 shuttles \\
\circuit{qugan_n39}      & timeout & 0 SWAP, 8 shuttles \\
\midrule
\circuit{qv_32}          & timeout & structural error \\
\circuit{knn_n25}        & placement-dep. & timeout (seed 0) \\
\circuit{vqe_uccsd_n28}  & scope-out & 0 SWAP, 1593 shuttles \\
\bottomrule
\end{tabular}
\end{table}

\textbf{Hub traps as an enabling capability.}
The \circuit{No-Hub} ablation supports a separate claim.
With hub traps disabled (the \circuit{No-Hub} and \circuit{Graphine(SWAP-only)} rows of Tab.~\ref{tab:method_matrix}), the compiler exceeds the 15-minute per-method budget on ten of the seventeen benchmarks (Tab.~\ref{tab:enabling}).
Of these ten, only \circuit{knn_n25} and \circuit{qv_32} fail to compile under \circuit{Proposed+Ring} within the same budget, and both are examined in Sec.~\ref{sec:evaluation:failure-modes}.
To check that this separation is not an artifact of the 15-minute budget, we re-ran the timed-out circuits under a two-hour per-method budget.
Setting aside \circuit{vqe_uccsd_n28} as the numerical scope-out case and \circuit{knn_n25} as a placement-dependent case treated in Sec.~\ref{sec:evaluation:failure-modes}, eight circuits remain.
On all eight, \circuit{No-Hub} returns no schedule at two hours, while \circuit{Proposed+Ring} compiles seven of them with zero SWAP gates in seconds to minutes.
The single exception is \circuit{qv_32}, which stops with a structural transport error rather than a timeout (Sec.~\ref{sec:evaluation:failure-modes}).
The effect is not confined to large circuits.
The nine-qubit \circuit{adder_9} exceeds even the two-hour \circuit{No-Hub} budget, whereas \circuit{Proposed+Ring} compiles it with zero SWAP gates and four shuttles.
Hub traps therefore convert routing problems that the SWAP-only formulation cannot close within a practical compile budget into problems that terminate in seconds to minutes.
The mechanism is search-space contraction.
A single shuttle through a hub replaces a SWAP chain whose length scales with the inter-qubit distance, which removes the combinatorial blow-up that the SWAP router faces on circuits with many long-range $\mathrm{CZ}$ pairs.
This second axis of contribution does not depend on the assumed shuttle fidelity (Sec.~\ref{sec:evaluation:sensitivities}).

\textbf{Excluded benchmark: \circuit{vqe_uccsd_n28}.}
The \circuit{vqe_uccsd_n28} benchmark exposes a numerical scope limitation of the metric, not a property of the compiler.
Its compiled schedule contains roughly $3.6\!\times\!10^{5}$ layers, and the layer-product fidelity estimate $F=\prod_\ell F_\ell$ underflows into the IEEE-754 subnormal range (\circuit{Proposed+Ring}: $F=2.47\!\times\!10^{-322}$; \circuit{Trivial-Layout-SWAP}: $F=1.28\!\times\!10^{-322}$).
The ratio $F_{\mathrm{PR}}/F_{\mathrm{TLS}}=1.92$ is well-defined arithmetically, but is numerically unreliable for ratios below the subnormal floor.
We exclude this benchmark from Fig.~\ref{fig:eval:fidelity_improvement}. Its compiled counts are listed in Tab.~\ref{tab:enabling} and the PTR ratio ($1.92\times$) is reported above.
A log-domain accumulator $\log F=\sum_\ell \log F_\ell$ would resolve the underflow at the cost of a one-line refactor of the fidelity estimator.
We defer this refactor and the corresponding sub-NISQ-regime study to future work.

\textbf{Failure-mode pointer.}
Two benchmarks (\circuit{knn_n25}, \circuit{qv_32}) are not completed by the proposed method within the 15-minute budget.
The failure modes differ.
\circuit{knn_n25} times out across all proposed variants and under the \circuit{No-Hub} ablation.
\circuit{qv_32} returns an explicit ``no valid transport'' transpiler error under the hub-enabled variants and times out under the SWAP-only ablation.
These cases are analyzed in Sec.~\ref{sec:evaluation:failure-modes}.

We close this subsection with a representative hub configuration.

\subsection{Sensitivities}
\label{sec:evaluation:sensitivities}

This subsection examines two modeling choices that affect the headline result.
The first is the per-shuttle fidelity.
The second is the parallel-shuttle assumption.
Across both sweeps, Regime~A retains its order-of-magnitude advantage over PTR.
Regime~B is sensitive to the assumed shuttle fidelity, as we report below.

\begin{figure}[t]
  \centering
  \includegraphics[width=\linewidth]{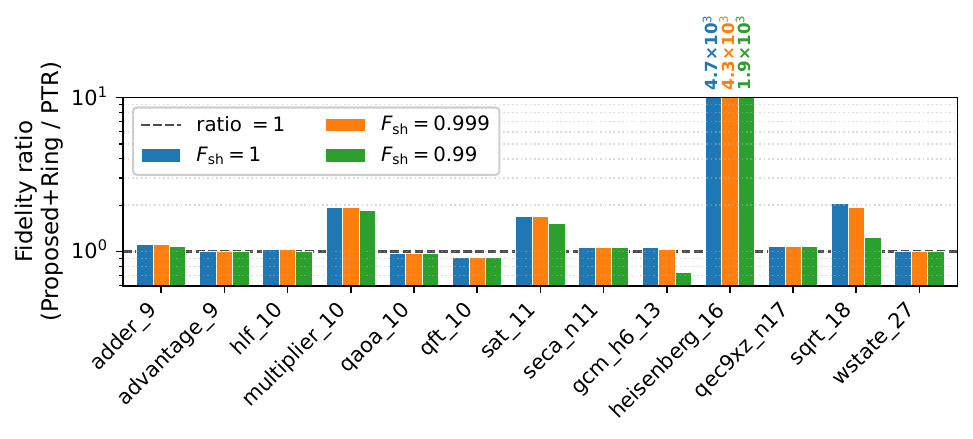}
\caption{Robustness of the \circuit{Proposed+Ring} advantage under a sweep of $F_{\mathrm{sh}}\in\{1.0,\,0.999,\,0.99\}$, with PTR as the baseline.
  Three grouped bars per benchmark correspond to the three $F_{\mathrm{sh}}$ values.
  PTR issues zero shuttles, so its fidelity is invariant under the sweep.
  Bars are capped at $10$; the \circuit{heisenberg_16} ratios, which exceed the cap at all three $F_{\mathrm{sh}}$ values, are printed above their bars.
  \circuit{qugan_n39} is omitted because PTR fails to compile it, leaving no ratio to plot.}
  \label{fig:eval:fs_sweep}
\end{figure}

\textbf{Shuttle-fidelity sensitivity.}
Our default operating point uses $F_{\mathrm{sh}}=1$ (Tab.~\ref{tab:hwparams}, footnote in Sec.~\ref{sec:evaluation:metrics}).
The layer-time $T_2$ dephasing during transport is captured separately, and the literature has not converged on a single per-shuttle infidelity for the AOD-mediated transfer model we assume \cite{bluvstein2024logical,wurtz2023aquila}.
To test how the headline result depends on this assumption, we sweep $F_{\mathrm{sh}}\in\{1.0,\,0.999,\,0.99\}$ and recompute the per-benchmark fidelity ratio against PTR (Fig.~\ref{fig:eval:fs_sweep}).
PTR issues zero shuttles, so its fidelity is invariant under the sweep.
The ratio bars therefore show how much of the \circuit{Proposed+Ring} advantage is consumed by per-shuttle fidelity loss.

In Regime~A, the advantage is preserved.
\circuit{heisenberg_16} (91 shuttles) drops from $4.73\!\times\!10^{3}$ at $F_{\mathrm{sh}}=1$ to $4.32\!\times\!10^{3}$ at $0.999$ and to $1.89\!\times\!10^{3}$ at $0.99$.
Three orders of magnitude of advantage remain at $F_{\mathrm{sh}}=0.99$.
\circuit{sqrt_18} (50 shuttles) drops from $2.04\!\times$ to $1.94\!\times$ to $1.24\!\times$, so the sign is preserved across the sweep.

In Regime~B, the picture is more nuanced.
\circuit{gcm_h6_13} (38 shuttles, $1.07\!\times$ at $F_{\mathrm{sh}}=1$) drops to $1.03\!\times$ at $0.999$ and to $0.73\!\times$ at $0.99$.
The ratio crosses unity.
The behavior is consistent with the cost balance.
At $F_{\mathrm{sh}}=0.99$, 38 shuttles introduce a multiplicative loss of about $0.68$, comparable to the gain from removing the SWAPs they replace.
The proposed method is a clear win in Regime~A and becomes a wash in Regime~B under pessimistic shuttle fidelity.
No benchmark in Regime~A crosses unity within the swept range.

\textbf{Parallel-shuttle sensitivity.}
Our headline execution-time numbers assume serial shuttle execution (model M0 in Sec.~\ref{sec:method:shuttle-models}).
The published hardware capability is a 2D rastering AOD that can move multiple atoms within a single transfer, subject to a non-intersection constraint on the source-destination segments \cite{bluvstein2024logical}.
We model this as M2: an offline post-processing pass that batches shuttle operations whose segments do not intersect and that lie within a contiguous window of layers in which no other operation acts on the shuttled atoms.

Across the seventeen benchmarks, the median M0-to-M2 reduction in total execution time is $0.0\%$ and the maximum is $22.0\%$, on \circuit{hlf_10} (4 shuttles batched into 3 AOD transfers).
\circuit{adder_9} reaches $17.3\%$, \circuit{sat_11} reaches $7.1\%$, \circuit{heisenberg_16} reaches $6.2\%$, and \circuit{gcm_h6_13} reaches $4.4\%$.
The remaining benchmarks fall below $4\%$.
The structural reason is the dominant routing pattern at NISQ scale on our placements, which is a shuttle followed by a $\mathrm{CZ}$ followed by a return shuttle.
Shuttled atoms are typically consumed by a $\mathrm{CZ}$ in the immediately following layer, which prevents batching with later shuttles whose source layers fall on the other side of the $\mathrm{CZ}$.
We therefore report M2 as a sensitivity result rather than a primary metric.
We expect the M0-to-M2 gap to widen at larger qubit counts where independent shuttle subtrees become more common, in line with the parallel speedups reported by hardware-side studies of fault-tolerant workloads \cite{bluvstein2024logical}.

\subsection{Failure modes}
\label{sec:evaluation:failure-modes}

We group the benchmarks by how routing succeeds or fails, which separates the role of hub traps from the limits of the proposed compiler.
Three groups emerge.

\textbf{Group 1: SWAP-only fails, hub traps recover.}
On these circuits, SWAP-only routing does not complete within the 15-minute budget, while the hub-enabled methods finish in seconds to minutes.
The group is not confined to large circuits.
\circuit{adder_9} has nine qubits, yet \circuit{No-Hub} and \circuit{Graphine(SWAP-only)} do not complete within the budget, whereas \circuit{Proposed} and \circuit{Proposed+Ring} compile it with zero SWAP gates and four shuttles.
The same pattern holds for larger members of the group such as \circuit{heisenberg_16} and \circuit{sqrt_18}.
The shared cause is the minimum-separation constraint.
Under this constraint the SWAP coupling graph is sparse, and the gate-based router explores long SWAP chains without reaching a feasible schedule in the allotted time.
Extending the per-method budget to two hours does not rescue any circuit in this group, as reported in Sec.~\ref{sec:evaluation:main}.
Hub traps add transit positions that shorten these chains, which is why the same circuits become tractable once hubs are available.
This group is the empirical basis for treating hub traps as an enabling capability rather than a fidelity optimization alone.

\textbf{Group 2: placement-dependent feasibility.}
\circuit{knn_n25} does not complete within the budget under the main placement (seed~0), and the \circuit{No-Hub} ablation times out on it as well.
Its status is not fixed, however.
We re-ran \circuit{knn_n25} under three additional placement seeds at a two-hour budget.
Under one seed the proposed pipeline compiles it with zero SWAP gates and seven shuttles, under another it stops with a structural no-valid-transport error, and under a third it does not return within the budget.
The same circuit therefore moves between completion, structural failure, and timeout as the placement changes, so its difficulty is a property of the layout rather than of the circuit alone.
We retain the seed~0 result in the main comparison and report \circuit{knn_n25} as a placement-sensitive instance rather than one the method cannot handle.

\textbf{Group 3: no valid transport.}
\circuit{qv_32} fails in a qualitatively different way.
The hub-enabled methods do not time out.
They stop with an explicit transpiler error reporting that a particular qubit pair has no valid transport, that the SWAP path is unavailable, and that shuttling and eviction are also unavailable.
The \circuit{No-Hub} ablation times out on the same circuit.
To test whether a different layout removes the closed pair, we re-ran \circuit{qv_32} under three additional placement seeds.
Every seed reproduces the same failure at a different qubit pair, so the closure is a property of the circuit under the minimum-separation constraint rather than of one particular layout.
We report this as an observed limit of the current placement and transport model, not as a proof that no schedule exists.
A transport model with additional move types might resolve the same pair.
Characterizing when transport is structurally unavailable, as opposed to merely expensive to find, is left to future work.

Groups 2 and 3 mark the boundary of the current method.
Group 1 is where the contribution of this work is concentrated.
The boundary cases clarify the scope of the enabling-capability claim.
Hub traps recover reachability that the SWAP-only formulation loses under the minimum-separation constraint, but they do not address a placement-dependent search cost that only a different layout resolves (Group~2), nor a placement-independent structural closure that no layout among those tried avoids (Group~3).
\section{Conclusion and Future Work}
\label{sec:conclusion}

We presented a two-step neutral-atom compilation workflow for monolithic single-zone devices: application-specific placement followed by hub-assisted routing with a per-gate SWAP-versus-shuttling decision.
Our main finding is that hub traps are an enabling capability.
Under the minimum-separation constraint, the SWAP-only configuration of our pipeline does not complete on a range of circuits, including circuits as small as nine qubits, within a practical time budget, and hub traps make these circuits compile in seconds to minutes while removing SWAP gates entirely.
The benefit is regime-dependent, concentrated on routing-dominated circuits and absent on routing-free ones, and it persists under pessimistic per-shuttle fidelity assumptions.
We also identify the boundary of the current method, where routing is either sensitive to the chosen placement or structurally closed across the placements we tried.

Several directions remain.
A quantitative predictor of the routing-dominated regime from circuit structure, such as the long-range interaction ratio, would turn the present qualitative separation into a usable criterion.
A theoretical characterization of when transport is structurally unavailable, as opposed to merely expensive to find, would strengthen the enabling-capability claim beyond the empirical evidence reported here.
Extending the evaluation to a larger benchmark suite, and to a direct, carefully matched comparison with movement-centric compilers under a common hardware model, is left to future work.

\section*{Data and Code Availability} Code and data will be made available upon publication.

% ===== Bibliography =====
\bibliographystyle{IEEEtran}
\bibliography{references}

\EOD

\end{document}